\documentclass[conference]{IEEEtran}
\IEEEoverridecommandlockouts
\usepackage{cite}
\usepackage{amsmath,amssymb,amsfonts}
\usepackage{algorithmic}
\usepackage{graphicx}
\usepackage{textcomp}
\usepackage{xcolor}
\usepackage[numbers,sort&compress]{natbib}
\usepackage{CJKutf8,multirow}
\def\BibTeX{{\rm B\kern-.05em{\sc i\kern-.025em b}\kern-.08em
    T\kern-.1667em\lower.7ex\hbox{E}\kern-.125emX}}
\begin{document}

\title{EffectiveASR: A Single-Step Non-Autoregressive Mandarin Speech Recognition Architecture with High Accuracy and Inference Speed\\
}

\author{\IEEEauthorblockN{1\textsuperscript{st} Ziyang Zhuang}
\IEEEauthorblockA{\textit{Ping An Technology} \\
}
\and
\IEEEauthorblockN{2\textsuperscript{nd} Chenfeng Miao}
\IEEEauthorblockA{\textit{Ping An Technology} \\
}
\and
\IEEEauthorblockN{3\textsuperscript{rd} Kun Zou}
\IEEEauthorblockA{\textit{Ping An Technology} \\
}
\and
\IEEEauthorblockN{4\textsuperscript{th} Ming Fang}
\IEEEauthorblockA{\textit{Ping An Technology} \\
}
\and
\IEEEauthorblockN{5\textsuperscript{th} Tao Wei}
\IEEEauthorblockA{\textit{Ping An Technology} \\
}
\and
\IEEEauthorblockN{6\textsuperscript{th} Zijian Li}
\IEEEauthorblockA{\textit{Georgia Institute of Technology} \\
}
\and
\IEEEauthorblockN{7\textsuperscript{th} Ning Cheng}
\IEEEauthorblockA{\textit{Ping An Technology} \\
}
\and
\IEEEauthorblockN{8\textsuperscript{th} Wei Hu}
\IEEEauthorblockA{\textit{Ping An Technology} \\
}
\and
\IEEEauthorblockN{9\textsuperscript{th} Shaojun Wang}
\IEEEauthorblockA{\textit{Ping An Technology} \\
}
\and
\IEEEauthorblockN{10\textsuperscript{th} Jing Xiao}
\IEEEauthorblockA{\textit{Ping An Technology} \\
}
}

\maketitle

\begin{abstract}
Non-autoregressive (NAR) automatic speech recognition (ASR) models predict tokens independently and simultaneously, bringing high inference speed. However, there is still a gap in the accuracy of the NAR models compared to the autoregressive (AR) models. In this paper, we propose a single-step NAR ASR architecture with high accuracy and inference speed, called EffectiveASR. It uses an Index Mapping Vector (IMV) based alignment generator to generate alignments during training, and an alignment predictor to learn the alignments for inference.
It can be trained end-to-end (E2E) with cross-entropy loss combined with alignment loss. The proposed EffectiveASR achieves competitive results on the AISHELL-1 and AISHELL-2 Mandarin benchmarks compared to the leading models. Specifically, it achieves character error rates (CER) of 4.26\%/4.62\% on the AISHELL-1 dev/test dataset, which outperforms the AR Conformer with about 30x inference speedup.
\end{abstract}


\section{Introduction}
\label{sec:intro}

In recent years, end-to-end (E2E) models have outperformed traditional hybrid systems in automatic speech recognition (ASR) tasks. Among the three widely used E2E methods—connectionist temporal classification (CTC) \cite{graves2006connectionist}, recurrent neural network transducer (RNN-T) \cite{graves2012sequence,graves2013speech}, and attention-based encoder-decoder (AED) \cite{chorowski2015attention,chan2016listen,miao2022towards}—AED models have become the preferred choice for sequence-to-sequence ASR modeling, thanks to their superior recognition accuracy, as seen in models like Transformer and Conformer. However, despite their strong performance, AED models utilize an auto-regressive (AR) decoder that generates tokens sequentially, with each token dependent on the preceding ones. This results in computational inefficiency, as decoding time increases linearly with the length of the output sequence. To address this issue and speed up inference, non-autoregressive (NAR) models have been introduced, allowing for parallel generation of output sequences.

Depending on the number of iterations in inference, NAR models are classified as iterative or single-step models. Among the former, A-FMLM \cite{chen2020non}, a non-autoregressive Transformer model (NAT), is the first attempt to introduce the conditional masked language model (CMLM) \cite{ghazvininejad2019mask} into ASR. A-FMLM is designed to predict masked tokens conditioned on unmasked ones and whole speech embeddings. However, its target token length needs to be pre-defined, which limits the model's performance. To address this issue, ST-NAT \cite{tian2020spike} uses a CTC module to predict the length of the target sequence. Unlike ST-NAT, Mask-CTC and its variants \cite{higuchi2020mask,higuchi2021improved,song2021non} propose to use the CMLM decoder to refine CTC. 

However, iterative models require multiple iterations to achieve a competitive result, limiting the speed of inference in practice. To overcome this limitation, single-step NAR models are proposed. LASO \cite{bai2021fast} implements parallel decoding of Transformer-style models based on the position-dependent summarizer (PDS) module, but it requires a predefined token length. InterCTC \cite{lee2021intermediate,nozaki2021relaxing} uses an additional intermediate loss of CTC to relax the conditional independence assumption of CTC models, improving the accuracy while maintaining the inference speed of the CTC model. CASS-NAT and its variants \cite{fan2021cass,fan2021improved,wang2022alignment,fan2023ctc} conduct in-depth exploration on the combination of CTC alignment and NAT, further improving the performance of the single-step NAR models. Different from the CTC-based or NAT-based NAR models mentioned above, Continuous Integrate-and-Fire (CIF) \cite{dong2020cif} is the first attempt to model ASR tasks by explicitly predicting the length of the target sequence. However, due to the conditional independence assumption, the accuracy of these mentioned NAR models is significantly inferior to the leading AR models.

Paraformer \cite{gao2022paraformer}, a  CIF-based model, designs a glancing language model (GLM) based sampler to strengthen the NAR decoder with the ability to model token interdependency. According to the Paraformer's report, it can achieve comparable performance to that of the AR transformer on open source Mandarin Chinese datasets \cite{bu2017aishell,du2018aishell}. A step further, E-Paraformer \cite{zou2024e-paraformer} proposes a novel monotonic alignment mechanism called Parallel Integrate-and-Fire (PIF), which not only enables parallel computation instead of the recursive mechanism in CIF, but also incorporates global context modeling capabilities and gets both better CER performance and faster speeds than the Paraformer. Despite the success of Paraformer and E-Paraformer in NAR ASR modeling, they both require a well-designed two-pass training process, which makes the model less compact and less efficient.

In this paper, we propose a single-step NAR ASR architecture inspired by the Hard Monotonic Alignment (HMA) method applied in \cite{miao2021efficienttts,zhuang2024improving}, termed EffectiveASR. The EffectiveASR is more compact compared to other high-performance NAR models, with efficient training and inference processes. And it achieves fairly competitive results on the public AISHELL-1 and AISHELL-2 benchmarks compared to the leading models.

\section{Methods}
\label{sec:methods}

\begin{figure}[htp]
\centering
\includegraphics[scale=0.6]{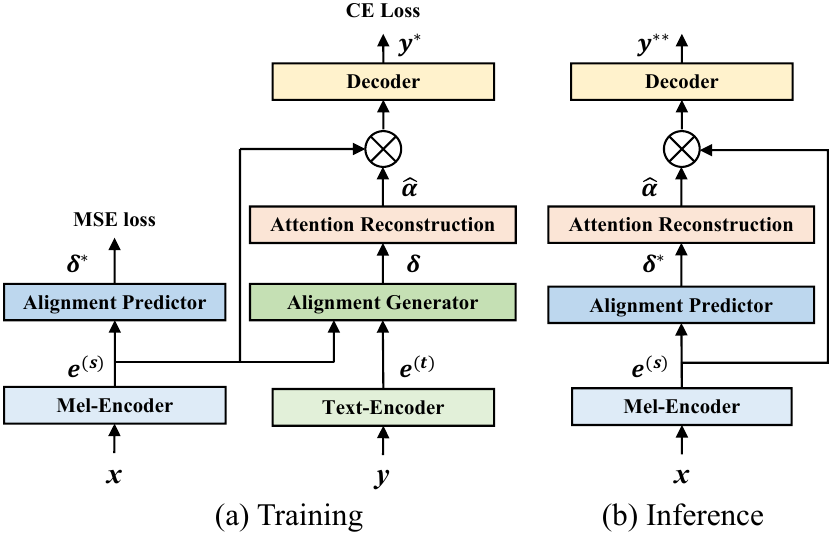}
\caption{The model architecture of the  proposed EffectiveASR. ($\bigotimes$ represents the matrix multiplication operation)}
\label{fig:model_arch}
\end{figure}

\subsection{Overview}
\label{sec:Overview}

The architecture of the proposed EffectiveASR is shown in Fig. \ref{fig:model_arch}. 
The training network consists of six modules, namely mel-encoder, text-encoder, alignment predictor, alignment generator, attention reconstruction, and decoder. The text-encoder and alignment generator are designed for training and are no longer activated during inference.

Let the acoustic feature sequence be $\boldsymbol x \in \mathcal{R}^T$ and the transcription sequence be $\boldsymbol y \in \mathcal{R}^L$, where $T$ and $L$ are the length of acoustic feature sequence and transcription sequence respectively. During training, the mel-encoder encodes $\boldsymbol x$ into the acoustic embeddings $\boldsymbol e^{(\mathrm{s})} \in \mathcal{R}^{T,d}$, and the text-encoder encodes $\boldsymbol y$ into text embeddings $\boldsymbol e^{(\mathrm {t})} \in \mathcal{R}^{L,d}$, where $d$ is the embedding dimension. The alignment generator leverages $\boldsymbol e^{(\mathrm{s})}$ and $\boldsymbol e^{(\mathrm{t})}$ to generate the alignment $\boldsymbol{\delta} \in \mathcal{R}^{T}$ between acoustic and text embeddings. $\boldsymbol{\delta}$ is then passed to the attention reconstruction module to construct an attention matrix $\boldsymbol{\hat{\alpha}} \in  \mathcal{R}^{L,T}$. By multiplying with $\boldsymbol{\hat{\alpha}}$, the acoustic embeddings $\boldsymbol e^{(\mathrm{s})}$ are transformed into semantic encodings, with a length corresponding to that of the output tokens. The semantic encodings are finally fed into the decoder to predict tokens $\boldsymbol {y}^{*} \in \mathcal{R}^L$. However, at the inference stage, the model cannot obtain alignments through the generator because of the lack of transcriptions. Therefore, during training, we use the predictor to learn the alignment produced by the generator. Then during inference, the output $\boldsymbol{\delta^*}$ of the alignment predictor is used to reconstruct the attention matrix and perform the inference of the model.

The mel-encoder is the same as the Conformer encoder \cite{gulati2020conformer}, consisting of several conformer blocks. The text-encoder and decoder are built with Transformer \cite{vaswani2017attention} encoder blocks. Other modules will be detailed in the following parts.

\begin{figure}[htp]
    \centering    
 \includegraphics[scale=0.9]{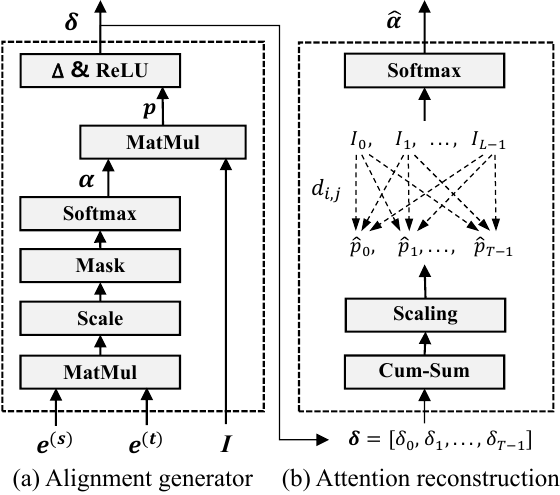}
    \caption{Visualization of alignment generator and attention reconstruction calculation process. (Operation $\Delta$ is defined as Eq. (\ref{delta_e}). Operation Cum-Sum is defined as Eq. (\ref{sum_e}).)}
    \label{fig:imv-ag}
\end{figure}

\subsection{Alignment generator}
\label{subsec:generator}

The alignment generator is built based on the Index Mapping Vector (IMV), proposed in \cite{miao2021efficienttts}, which provides a creative way to describe the alignment between input and output sequences. The calculation process of IMV-based alignment for ASR is depicted in Fig. \ref{fig:imv-ag}(a). First, a scaled dot-product attention matrix $\boldsymbol{\alpha} \in \mathcal{R}^{T,L}$ is built by Eq. (\ref{dot-product-att}), with $\boldsymbol e^{(\mathrm{s})}$ as Query and $\boldsymbol e^{(\mathrm{t})}$ as Key:
\begin{equation}
\label{dot-product-att}
    \alpha_{i,j} = \frac{\mathrm{exp}((\boldsymbol e_{i}^{(\mathrm{s})} \cdot \boldsymbol e_{j}^{(\mathrm{t})})*d^{-0.5})}{\sum_{j=0}^{L-1} \mathrm{exp}((\boldsymbol e_{i}^{(\mathrm{s})} \cdot \boldsymbol e_{j}^{(\mathrm{t})})*d^{-0.5})}
\end{equation}
where $0 \le i \le T-1$, $0 \le j \le L-1$, and $d$ is the dimension of $\boldsymbol e^{(\mathrm{s})}$ and $\boldsymbol e^{(\mathrm{t})}$. 
Then the IMV $\boldsymbol p$ is calculated by Eq. (\ref{e_dummy}):
\begin{equation}
    \boldsymbol p = \boldsymbol \alpha \cdot \boldsymbol I
    \label{e_dummy}
\end{equation}
where $\boldsymbol I=[0,1,..., L-1]$ is the index vector of transcription, and $\Sigma_{j=0}^{L-1}\alpha_{i,j}=1$. Here $p_i=\Sigma_{j=0}^{L-1}(\alpha_{i,j}*j)$ is the weighted sum of the attention weight between the $i$-th input frame and each output token and the position index of each output token. $p_i$ can be seen as the expected alignment location of the $i$-th input frame in the length range of the output text. 
We define the increment of alignment locations of adjacent audio frames in the transcription as ${\Delta p_i}$, which is calculated by Eq. (\ref{delta_e}).
According to the monotonicity assumption in \cite{miao2021efficienttts,zhuang2024improving,miao2022towards}, when there is a monotonic correspondence between audio and transcription, it can be deduced that $\Delta p_i \ge 0$, and the proof process can be found in \cite{miao2021efficienttts,miao2022towards}.
\begin{align}
\label{delta_e}
\Delta p_i=&p_i-p_{i-1},&1\le i \le T-1    
\end{align}

However, at the beginning of training, the model has not been trained well enough to satisfy $\Delta p_i \ge 0$ easily. To guarantee a strictly monotonic constraint, $\Delta p_i$ is activated by ReLU, as shown in Eq. (\ref{relu_e}):
\begin{align}
\label{relu_e}
 \delta_{i}=\left\{\begin{array}{ll}
0, & i=0 \\
{\rm ReLU}(\Delta p_i), &1\le i \le T-1
\end{array}\right.
\end{align}
where $\delta_0$ is set to $0$. $\delta_i$ can be seen as the weight of each encoder step involved in the calculation of decoding label.

\begin{table*}[htp]
\centering
\caption{CER(\%) and RTF results of our models on AISHELL-1 and AISHELL-2 tasks and the comparison with previous AR and NAR works. (\dag : RTF is evaluated with batchsize of 8, \ddag : Model parameters in the inference phase.)}

\scalebox{1.5}{
\begin{tabular}{llllllll}
\hline
\multirow{2}{*}{}   & \multirow{2}{*}{Model} & \multirow{2}{*}{LM} & \multicolumn{2}{l}{AISHELL-1} & {AISHELL-2} & \multirow{2}{*}{Params} & \multirow{2}{*}{RTF ($\downarrow$)} \\ \cline{4-6}
\multicolumn{3}{l}{ } & {dev}    & {test}   & {\hfil test-ios}    \\ \hline \hline
\multirow{2}{*}{AR}
& Transformer \cite{bai2021fast} &        w/o         &  6.1    &   6.6      &     {\hfil 7.1}          &            67.5M           &           0.1900           \\
& Conformer \cite{gao2023funasr} &        w/o         &  -    &   5.21  &            {\hfil 5.83}          &         46.25M          &           0.2100           \\ \hline
\multirow{9}{*}{NAR} 
& A-FMLM \cite{chen2020non}  &      w/o               &       6.2      &      6.7          &    {\hfil -}           &       -          &         0.2800               \\
&  CTC-enhanced \cite{song2021non}  &      w/o    &       5.3      &      5.9          &    {\hfil 7.1}     &       29.7M          &         0.0037 $\dag$ \\
& LASO-big \cite{bai2021fast}  &      w/o               &       5.9      &      6.6          &    {\hfil 6.7}           &       80.0M     &         0.0040    \\

&  TSNAT \cite{2021TSNAT}  &      w/o    &       5.1      &      5.6         &      {\hfil -}     &       87M          &         0.0185 \\
&  Improved CASS-NAT \cite{fan2021improved}  &      w/o    &       4.9      &      5.4          &    {\hfil -}     &       38.3M         &         0.0230 \\
&    AL-NAT \cite{wang2022alignment}     &      w/o               &       4.9      &      5.3        &    {\hfil -}           &       71.3M          &         0.0050               \\ 

&    Paraformer \cite{gao2022paraformer}     &      w/o               &       4.6      &      5.2       &    {\hfil 6.19}           &       46.2M          &         0.0168               \\
&    E-Paraformer \cite{zou2024e-paraformer}     &      w/o               &       4.36      &      4.79       &    {\hfil 6.08}           &       43.6M          &         0.0069               \\
&   \textbf{EffectiveASR Base (Ours)}    &           w/o               &    4.30     &    4.66       &       {\hfil 6.03}       &    43.6M \ddag                    &             0.0058         \\
 &   \textbf{EffectiveASR Large (Ours)}  &           w/o   &    \textbf{4.26}    &   \textbf{4.62}    &    {\hfil \textbf{5.76}}       &   76.0M \ddag         &          0.0066            \\ \hline
\end{tabular}
}
\label{results-aishell}
\end{table*}

\subsection{Attention reconstruction}
\label{subsec:reconstruction}

After obtaining alignment $\boldsymbol \delta$, our goal is to reconstruct an attention matrix, such that the acoustic embeddings $\boldsymbol e^{(\mathrm{s})}$ of length $T$ can be transformed into semantic encodings of length $L$ by multiplying with the attention matrix. We follow the similar idea of \cite{miao2021efficienttts,zhuang2024improving} in reconstructing the attention matrix and the construction process of the attention matrix can be summarized as shown in Fig. \ref{fig:imv-ag}(b).
First, a monotonically increasing alignment position vector $\boldsymbol{p \mathrm{'}}$ is generated by accumulating $\delta_i$, as shown in Eq. (\ref{sum_e}). 
Next, a scaling strategy, as described in Eq. (\ref{scale_e}), is applied to ensure that the maximum value of $\boldsymbol{p'}$ matches the length of the output tokens.
\begin{align}
\label{sum_e}
p_i{'}=&\sum_{m=0}^{i} \delta_{m},&0\le i \le T-1
\end{align}
\begin{align}
\label{scale_e}
\hat{p}_i=&\frac{p_i{'}-p_0{'}}{p_{T-1}{'}-p_0{'}} * (L-1),&0\le i \le T-1
\end{align}

Since both $\boldsymbol{\hat{p}}$ and transcription index vector $\boldsymbol{I}$ are monotonic sequences ranging from $0$ to $L-1$, a distance-aware attention reconstruction is performed by Eq. (\ref{gaussian_att}):
\begin{equation}
\hat{\alpha}_{i,j} = \frac{{\rm exp}(-\sigma^{-2}*d_{i,j})}{\sum_{i=0}^{T-1} {\rm exp}(-\sigma^{-2}*d_{i,j})}
    \label{gaussian_att}
\end{equation}
where $d_{i,j} = (\hat{p}_i - I_j)^2, 0 \le i \le T-1, 0 \le j \le L-1$. $d_{i,j}$ represents the semantic distance between the $i$-th encoder frame and the $j$-th token. $\sigma$ is a learnable hyperparameter. It's evident that as the distance $d_{i,j}$ decreases, the attention weight $\hat{\alpha}_{i,j}$ increases.

\subsection{Alignment predictor}
\label{subsec:predictor}

As mentioned in Section \ref{sec:Overview}, an alignment predictor is required to generate alignments at inference stage. In EffectiveASR, the alignment predictor is built with two 1D-Convolution layers, each followed by layer normalization and ReLU function. During training, it takes acoustic embeddings $\boldsymbol e^{(\mathrm{s})}$ as inputs to generate a predicted alignment $\boldsymbol {\delta}^{*}$, while using the alignment $\boldsymbol {\delta}$ produced by the alignment generator as the label to update the predictor. In the inference stage, we use the alignment $\boldsymbol {\delta}^{*}$ produced by the alignment predictor to construct the attention matrix, as shown in the Fig. \ref{fig:model_arch}(b). At the same time, we accumulate the $\boldsymbol {\delta}^{*}$ and round it to predict the length of output tokens.
The Mean Square Error (MSE) loss is used to learn the alignment, and the total loss is defined as:
\begin{equation}
    \mathcal{L}_{total} = \mathcal{L}_{\rm{CE}}(\boldsymbol{y},\boldsymbol{y^*}) + \lambda \mathcal{L}_{\rm {MSE}}(\boldsymbol{\delta}, \boldsymbol{\delta^*})  \\
\label{loss}
\end{equation}
where $\lambda$ is the weight of alignment loss, and $\mathcal{L}_{\rm{CE}}$ is the cross-entropy (CE) loss between the text outputs and labels.

\section{Experiments}
\label{sec:experiments}
\subsection{Experimental Setup}
\label{subsec:setup}
The proposed method is evaluated on two public Mandarin corpus: 178 hours AISHELL-1 \cite{bu2017aishell} and 1000 hours AISHELL-2 \cite{du2018aishell}. We use 80-channel filter banks computed from a 25 ms window with a stride of 10 ms as features. SpecAugment \cite{park2019specaugment} and speed perturbation are used for data augmentation for all experiments. The sizes of the vocabulary for the AISHELL-1 and AISHELL-2 tasks are $4,233$ and $5,208$, respectively.
The EffectiveASR model is built in base and large size. For both sizes, the layer numbers of \{mel-encoder, text-encoder, alignment predictor, decoder\} are set to \{12, 1, 2, 6\}, while the attention heads number of base model is $4$ with hidden dimension of $256$ and the attention heads number of large model is $6$ with hidden dimension of $384$. The hyperparameter $\lambda$ in Eq. (\ref{loss}) is set to $1.0$ and the learnable hyperparameter $\sigma$ in Eq. (\ref{gaussian_att}) is initialized with $0.5$. The proposed models are developed by ESPnet \cite{watanabe2018espnet} \footnote{https://github.com/espnet/espnet.git}, which is a popular open-source ASR toolkit. All experiments are performed on $8$ NVIDIA Tesla V100 GPUs, following the default configurations of ESPnet.

We use the character error rate (CER) to evaluate the performance of different models and the real-time factor (RTF) to measure the inference speed. 

\subsection{Results}
\label{subsec:results}

The evaluation results of AISHELL-1 and AISHELL-2 are detailed in Table \ref{results-aishell}. The CER and RTF values of previous works are obtained from the papers' report, except that results of the ESPnet Conformer are obtained from the ESPnet ofﬁcial repository. The RTF of our models is calculated on the AISHELL-1 test set with a batch size of $1$. For a fair comparison with the published work, none of our experiments in Table \ref{results-aishell} uses an external language model (LM) or pre-training.

For AISHELL-1 task, the proposed EffectiveASR models (both base and large size) outperform all other models presented. On the AISEHLL-1 test set, our large model achieves a CER of 4.62\%,  which is 0.58\%/0.17\% (absolutely) better than the previous leading results of NAR models (from Paraformer \cite{gao2023funasr}/E-Paraformer \cite{zou2024e-paraformer}). Furthermore, the inference speed of our large model is twice more than that of Paraformer. With the application of PIF, E-Paraformer also outperforms Paraformer in both inference speed and CER. However, the two-pass training process of E-Paraformer makes the model less compact and less efficient compared to the single-step architecture of EffectiveASR. When compared with the AR models, our model not only outperforms AR Conformer in CER, but also has inference speed 30x faster than the AR Conformer, which is an impressive performance.

Since AISHELL-1 contains less than 200 hours traning data, we also validate our method on a larger corpus, 1000 hours AISHELL-2. On the AISHELL-2 test ios set, our large model achieves a CER of 5.76\%, which also outperforms all AR and NAR models presented.

\begin{figure}[htp]
    \centering   
 \includegraphics[scale=0.6]{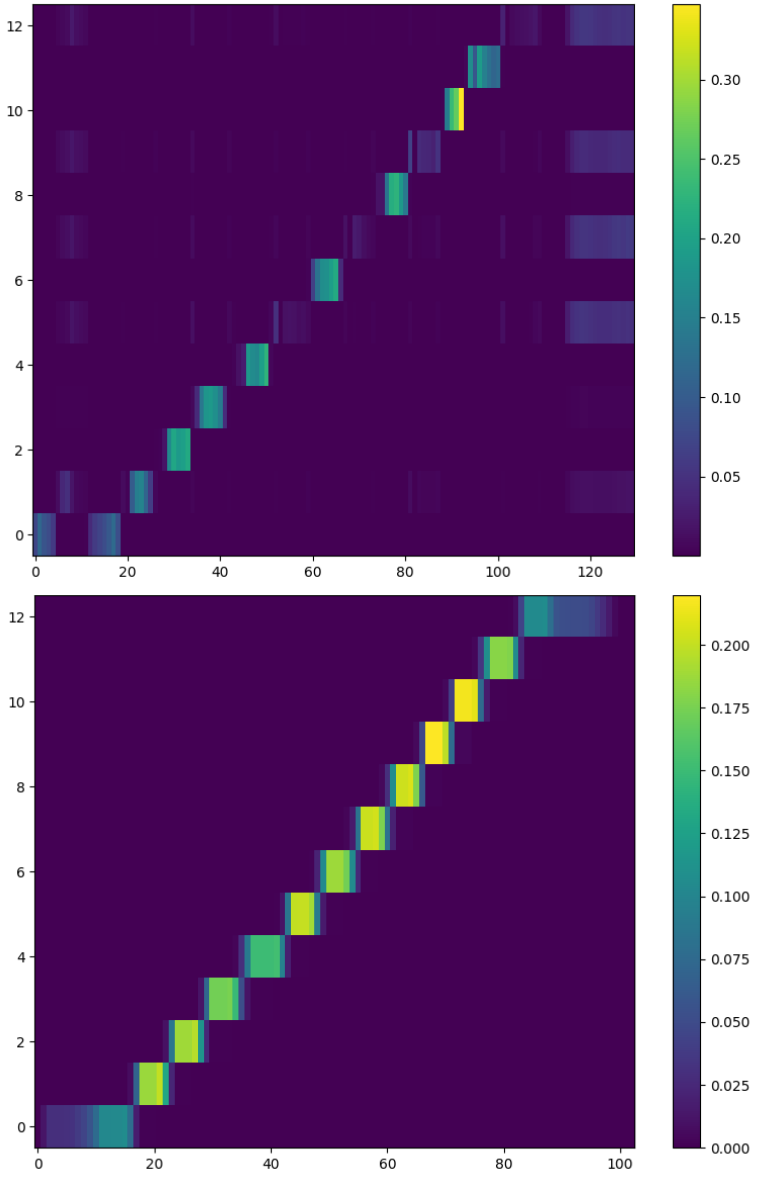}
 \caption{Alignment plots of both the baseline and proposed model. The first plot is the alignment plot of the Conformer, while the second plot is the alignment constructed by the proposed EffectiveASR. The horizontal axis represents the input frame step, and the vertical axis represents the output step.}
    \label{fig:align}
\end{figure}

To further analyze the behavior of the proposed model, we plot the alignment matrices in Figure \ref{fig:align}, comparing the Conformer with the proposed EffectiveASR. The alignment matrix of the Conformer is taken from the model's final cross-attention layer. As depicted in the figure, the proposed approach yields more consistent and refined alignments compared to the Conformer model, thereby contributing to an enhanced CER performance.

\section{Conclusion}
\label{sec:conclusion}

In this paper, we propose a novel single-step NAR ASR architecture called EffectiveASR. By leveraging our proposed methods for generating monotonic alignments and constructing alignment matrices, EffectiveASR not only possesses a simpler structure but also achieves leading performance in terms of inference efficiency and accuracy. Experiments on open-source Mandarin datasets demonstrate that the proposed EffectiveASR achieves better CER performance to the leading AR Conformer, with a 30x decoding speedup. In the future, we will focus on exploring the application of EffectiveASR in English speech recognition.

\vfill\pagebreak

\bibliographystyle{IEEEtran}
\bibliography{mybib}


\end{document}